\documentclass[12pt]{iopart}
\usepackage[dvips]{graphicx}

\usepackage{braket}
\usepackage{amsfonts}
\usepackage[colorlinks=true]{hyperref}
\usepackage{hyperref}
\usepackage{xcolor}

\usepackage{booktabs}

\usepackage{aas_macros}

\usepackage{caption}
\usepackage{graphicx}
\usepackage{subcaption}

\usepackage{bm}

\def\aap{\textit{Astronomy and Astrophysics}}

\def\be{\begin{equation}}
\def\ee{\end{equation}}
\def\bea{\begin{eqnarray}}
\def\eea{\end{eqnarray}}

\def\L{\mathcal{L}}

\begin{document}
\title[De Sitter space-times in Entangled Relativity]{De Sitter space-times in Entangled Relativity}

\author{Olivier Minazzoli}
\ead{ominazzoli@gmail.com}
\address{Artemis, Universit\'e C\^ote d'Azur, CNRS, Observatoire C\^ote d'Azur, BP4229, 06304, Nice Cedex 4, France}
\begin{abstract}
It is argued that de Sitter space-times might be solutions of entangled relativity once the quantum trace anomaly from matter fields in curved space-times is taken into account. This hypothesis would be an elegant solution to the acceleration of the expansion of the universe within the rigid framework of entangled relativity. 
\end{abstract}
\maketitle

\section{Introduction}

These days, evidence indicates that the universe today is very well approximated by a de Sitter universe, as it seems that a cosmological constant---or something that behaves like it---indeed started to lead the dynamics of the universe around some 4 billion years ago \cite{ryden:2016bk}. 

Entangled relativity has been suggested as a potential solution to the fact that vacuum solutions are allowed in general relativity, whereas, arguably, it violates Mach's Principle of Einstein \cite{einstein:1918an}\footnote{A translation in English of the original paper in German is available online at \href{https://einsteinpapers.press.princeton.edu/vol7-trans/49}{https://einsteinpapers.press.princeton.edu/vol7-trans/49}.} \cite{norton:1995cf,hoefer:1995cf,pais:1982bk}. Indeed, the action of the theory involves a pure multiplicative coupling between the matter Lagrangian and the Ricci scalar \cite{ludwig:2015pl,minazzoli:2018pr}, such that one fundamentally has to consider both gravity and matter at the same time.

Unlike in general relativity, it is not possible to add a cosmological constant to the action of entangled relativity \cite{ludwig:2015pl,minazzoli:2018pr} without spoiling the theory \cite{minazzoli:2018pr}. Hence, the fact that the universe today is well approximated by a de Sitter universe may be seen as a serious challenge for the theory of entangled relativity \cite{minazzoli:2018pr}---especially given the fact that, unlike general relativity, entangled relativity is free of any parameter at the classical level. In \cite{minazzoli:2018pr} was proposed a far-fetched hypothesis on how entangled relativity might still be able to lead to a de Sitter universe at late times: it was hypothesized that quantum corrections to the action of the theory may have the very precise properties that would lead to a dynamics that would be equivalent to the one produced by the cosmological constant in general relativity. In the present manuscript, I propose a much simpler, plausible and elegant hypothesis, namely, that the acceleration of the expansion of the universe may be driven by the quantum trace anomaly of matter fields in the cosmological space-time. 

The basics of the the reasoning is that because entangled relativity reduces to---or tends to---general relativity without a cosmological constant whenever $\L_m = T$ on-shell---where $T$ is the trace of the stress-energy tensor---the explanation of the acceleration of the expansion of the universe through the trace anomaly that has been proposed in the context of general relativity \cite{schutzhold:2002pl} might work for entangled relativity as well, simply because the phenomenology of the theory seems to remain equivalent to the one of general relativity in that case. 

The impossible existence of gravity without matter, and vice versa, in entangled relativity is obvious from its action \cite{ludwig:2015pl,minazzoli:2018pr}:
\be
\label{eq:ER}
S=-\frac{\xi}{2c} \int \mathrm{d}^{4} x \sqrt{-g} \frac{\mathcal{L}_{m}^{2}}{R},
\ee 
where the constant $\xi$ has the dimension of the usual coupling constant of general relativity $\kappa \equiv 8 \pi G / c^{4}$---where $G$ is the Newtonian constant and $c$ the speed of light. It comes from the fact that one has replaced the usual additive coupling between matter and geometry by a pure multiplicative coupling. $\xi$ defines a novel fundamental scale that is relevant at the quantum level only---because it does not appear in the field equations that derive from the action (\ref{eq:ER}) \cite{minazzoli:2018pr}. 
It turns out that entangled relativity can be rewritten---and is more easily understood---in a dilaton equivalent\footnote{This dilaton theory is equivalent, at least at the classical level, as long as 
$\mathcal{L}_m/R <0$. Notably, it seems that one must always consider cases such that $(R,\mathcal{L}_m) \neq 0$ when one uses the dilaton form of entangled relativity, although $R$ and $\mathcal{L}_m$ can be arbitrarily small in principle.} form that reads \cite{ludwig:2015pl,minazzoli:2018pr}
\be
\label{eq:sfaction}
S=\frac{1}{c} \frac{\xi}{\tilde \kappa} \int d^{4} x \sqrt{-g}\left[\frac{\phi R}{2 \tilde \kappa}+\sqrt{\phi} \mathcal{L}_{m}\right],
\ee
where $\tilde \kappa$ is an effective coupling constant between matter and geometry, with the dimension of $\kappa$. In particular, the quantity $\sqrt{\phi} = - \tilde \kappa \L_m/R$ behaves as an additional massless scalar degree of freedom with respect to general relativity. The equivalence between the two actions in Eqs. (\ref{eq:ER}) and (\ref{eq:sfaction}) is similar to the equivalence between $f(R)$ theories and the corresponding specific scalar-tensor theories \cite{capozziello:2015sc}. The corresponding field equations read 
\bea
&&G_{\alpha \beta}=\tilde \kappa \frac{T_{\alpha \beta}}{\sqrt{\phi}}+\frac{1}{\phi}\left[\nabla_{\alpha} \nabla_{\beta}-g_{\alpha \beta} \square\right] \phi,\label{eq:metric} \\
&&\frac{3}{\phi} \square \phi=\frac{\tilde \kappa}{\sqrt{\phi}}\left(T-\mathcal{L}_{m}\right), \label{eq:sceq}
\eea
where $G_{\alpha \beta}$ is the Einstein tensor. The conservation equation reads
\be
\label{eq:noncons}
\nabla_{\sigma}\left(\sqrt{\phi} T^{\alpha \sigma}\right)=\mathcal{L}_{m} \nabla^{\alpha} \sqrt{\phi},
\ee
with
\be
T_{\mu \nu} \equiv-\frac{2}{\sqrt{-g}} \frac{\delta\left(\sqrt{-g} \mathcal{L}_{m}\right)}{\delta g^{\mu \nu}}.
\ee
$\tilde \kappa$ takes its value from the asymptotic cosmological behavior of the effective scalar degree of freedom \cite{minazzoli:2018pr}, as well as the considered normalization of $\phi$. For instance, for the normalization $\phi_{today} =1$, $\tilde \kappa = 8 \pi G/c^4$, where $G$ is the measured value of Newton's constant today \cite{minazzoli:2013pr}.
From this alternative action, one can easily see why entangled relativity reduces to general relativity when the variation of the scalar field degree of freedom vanishes. 

\section{Argument}

\subsection{Generic conditions that freeze the scalar-field}

In this framework, the scalar degree of freedom is not---or weakly---sourced for any matter field that is such that $\L_m \sim T$ on-shell\footnote{By \textit{on-shell}, here I mean the effective value that takes the Lagrangian when the matter field solutions are injected into its formal definition. In particular, the value of $\mathcal{L}_m$ that appears in Eq. (\ref{eq:sceq}).}, such that the phenomenology of the theory is very close to the one of general relativity in all such situations---such as in the solar system \cite{minazzoli:2013pr} or more generally in the matter era \cite{minazzoli:2014pr}; while it may differ when $\L_m \not \sim T$, such  as for neutron stars for instance \cite{arruga:2021pr}. This is due to the cancellation in the right-hand-side of equation (\ref{eq:sceq}) for $\L_m = T$ on-shell, which has been named \textit{intrinsic decoupling} in \cite{minazzoli:2013pr,minazzoli:2016pr}. 

In particular, the cosmological dynamics of this type of theories---i.e. like Eq. (\ref{eq:sfaction})---has been studied, notably in \cite{minazzoli:2014pr, minazzoli:2014pl}. In particular, it has been shown that, assuming a flat Friedmann-Lema\^itre-Robertson-Walker (FLRW) metric, the phenomenology of any theory which action is
\be
S\propto \int d^{4} x \sqrt{-g}\left[\frac{1}{2 \tilde \kappa}\left(\phi R+\frac{\omega(\phi)}{\phi} (\partial \phi)^2 \right)+ \sqrt{\phi}\mathcal{L}_{m}\right],
\ee
converges toward the one of general relativity without a cosmological constant during the so-called \textit{matter era}. This becomes even more obvious with the special case given in Eq. (\ref{eq:sfaction})---that is, for $\omega(\phi) =0$.\footnote{Note that the relative deviation from general relativity in the Solar System, which is of the order of the ratio between the pressure and the energy density of the source of the gravitational field \cite{minazzoli:2013pr}, is about $10^{-10}$ for the Earth, and not $10^{-6}$ as was written in \cite{minazzoli:2013pr}. As a consequence, $\omega \sim 0$ will not be constrained by forthcoming measurement of the gravitational redshift around the Earth, for instance with \cite{cacciapuoti:2007np}---contrary to what was written in \cite{minazzoli:2013pr}.} Indeed, for an energy momentum tensor that is dominated by a dust field, one has $\L_m \sim -\rho_{dust}=T$, where $\rho_{dust}$ is the energy density of the free point particles. Hence, the right-hand-side of Eq. (\ref{eq:sceq}) cancels out. (Note that an electromagnetic pure radiation field would not act as a source of the dilaton field either, because in that case one has $\L_{radiation} \propto E^2 - B^2 = 0$ and $T=0$, such that $\L_m -T =0$ in Eq. (\ref{eq:sceq}) as well). Assuming a flat FLRW metric, the dilaton field equation (\ref{eq:sceq}) for any field that is such that $\L_m \sim T$---such as for the energy momentum tensor that is dominated by a dust field---simply reduces to
\be
\ddot \phi + 3 H \dot \phi \sim 0, \label{eq:friction}
\ee
where $H$ is the usual Hubble parameter constructed upon the FLRW conformal factor $a$ with $H=\dot a/a$, such that the second term on the left-hand-side of equation (\ref{eq:friction}) acts as a friction term that freezes the value of the scalar-field as long as $H>0$.

\subsection{Assumption}

The hypothesis I consider in the present manuscript is that the dilaton field directly couples to the quantum trace anomaly of the standard model (SM) in the effective action, such that the on-shell Lagrangian gets a contribution from the standard model trace anomaly. Formally, the assumption reads
\be
\L_{anomaly} = T_{anomaly}^{SM},\label{eq:assum}
\ee
in Eq. (\ref{eq:sceq}), such that, overall, one has in Eq. (\ref{eq:sceq})
\be
\mathcal{L}_m = T_{dust} + T_{radiation} + T_{anomaly}^{SM} + \L_{other}, \label{eq:sumL}
\ee
where $T_{dust} = -\rho_{dust}=\L_{dust}$, $T_{radiation}= 0=\L_{radiation} \propto E^2-B^2 $ and $\L_{other}$ are other types of contributions that may not reduce to the trace of their stress-energy tensor---such as magnetic fields, which are such that $\L_{magnetic} \propto B^2 \neq T$ \cite{thorsrud:2012jh}.
In a nutshell, provided that $\L_{other} \sim 0$, Eq. (\ref{eq:assum}) implies that $\L_m \sim T$ in both the matter and dark energy eras, such that one has Eq. (\ref{eq:friction}) in both eras, and the scalar field $\phi$ becomes a constant until the end of time. As a consequence, all the field equations converge to the ones of general relativity asymptotically, with the convergence starting at least at the begining of the matter era.\footnote{What happens during the radiation era has yet to be studied, and will depend on the on-shell value of $\L_m$ during the various phases of this era.}

Although the matter Lagrangian in the original action of entangled relativity (\ref{eq:ER}) might not simply be the standard model of particles, it must reduce (at least approximately) to the standard model Lagrangian at late times---simply because it is what seems to be consistent with observations.

\subsection{The equivalence principle}

The assumption in Eq. (\ref{eq:assum}) is motivated by the \textit{universality of free fall}---or, more precisely, by the \textit{weak equivalence principle} \cite{will:2014lr}---which seems to be a correct property of nature to a very good level of accuracy \cite{touboul:2019cq}. Because, according to current knowledge \cite{donoghue:2014bk}, the mass of nucleons comes from the quantum field theory trace anomaly, which includes a classical contribution of fermions to the overall mass, in addition to a purely quantum contribution \cite{damour:2010pr,nitti:2012pr}. Hence, if the scalar field degree of freedom couples identically to all the terms of the quantum trace anomaly, one would not have a violation of the \textit{weak equivalence principle} \cite{minazzoli:2016pr,nitti:2012pr}. If the assumption in Eq. (\ref{eq:assum}) does not hold however, not only the proposal presented in this manuscript would not work, but the theory may not be able to explain the manifest \textit{weak equivalence principle} at the required level of sensitivity either \cite{minazzoli:2016pr,nitti:2012pr}.

Indeed, the main thing to keep in mind in entangled relativity is that a variation of the dilaton field $\phi$ induces in effect a variation of all the effective coupling constants---whether it is the dimensionfull effective constant of Newton, or the dimensionless coupling constants of the standard model particle sector \cite{hees:2014pr}. But the constraints on the variations of these constants at the cosmological scales are pretty strong already, see e.g. \cite{hees:2014pr,minazzoli:2014pr,holanda:2016pr} and references therein, suggesting that the scalar field is constant---or close to be---at late times. Therefore, this suggests that whatever is the source of the scalar-field, it must be such that the on-shell Lagrangian is equal to the trace of the stress-energy tensor, such that the scalar-field is unsourced in Eq. (\ref{eq:sceq}). Indeed, due to the friction term in the FLRW d'Alembertian, the scalar-field converges to a---or would remain---constant in that case.
As mentioned above, this is already what happens for dust fields. Hence, if the assumption in Eq. (\ref{eq:assum}) is correct, and if it leads to a de Sitter expansion as we will see below, not only would the theory satisfy the \textit{weak equivalence principle}, but it would also explain other manifestations of the \textit{equivalence principle} at the cosmological level\footnote{See, e.g., \cite{hees:2014pr}.} during the so-called \textit{matter} and \textit{dark energy eras}.

\subsection{Trace anomaly of self-interacting quantum fields in a curved space-time}

Note, however, that in the case of the study of the  \textit{weak equivalence principle} \cite{minazzoli:2016pr, nitti:2012pr}, one is dealing with the trace anomaly of self-interacting quantum fields in a space-time that can be approximated to be flat; whereas in the present manuscript, the trace anomaly corresponds to self-interacting quantum fields in a curved space-time.

Investigating the trace anomaly of self-interacting quantum fields in a curved cosmological space-time is a difficult task, but an estimate has been put forward for quantum chromodynamics (QCD) some time ago in \cite{schutzhold:2002pl}, and it indicates that such a trace-anomaly should be of the order of $H \Lambda_\textrm{QCD}^3$---or close to be \cite{holdom:2011pl}---where $H$ is the Hubble parameter and $\Lambda_\textrm{QCD}$ is the QCD chiral symmetry breaking scale. It turns out that $H_0 \Lambda_\textrm{QCD}^3$ is remarkably of the order of magnitude of the inferred value of the cosmological constant $\Lambda$, despite the two very different scales involved \cite{schutzhold:2002pl}---where $H_0$ is the Hubble constant. 

Let us note, otherwise, that the trace anomaly from early-universe quantum matter fields has also been investigated as a potential source of the inflation \cite{starobinsky:1980pl,hawking:2001pr,bamba:2014pr}---also known as \textit{anomaly-induced inflation} \cite{netto:2016ej}---and that this scenario is still consistent with observations of the cosmological microwave background to date \cite{planck:2020in}.

Whether or not the explanation for the late-times acceleration of the expansion of the universe proposed in this manuscript could be extended to encompass an explanation of an early-times acceleration of the expansion of the universe (inflation) has to be studied.

\subsection{Asymptotic behavior}

In what follows, we assume that $\L_{other}$ can be treated has a perturbation such that $\L_{other}=0$ in Eq. (\ref{eq:sumL}) at leading order. 

Because the dilaton field freezes and remains constant for $\mathcal{L}_m = T$ in Eq. (\ref{eq:sceq}) for a FLRW universe that is such that $H>0$, the asymptotic behavior of entangled relativity when taking into account the quantum trace anomaly in Eq. (\ref{eq:sumL}) is the same as the one of general relativity when taking into account the contribution from the quantum trace anomaly, as one can see from Eq. (\ref{eq:metric}--\ref{eq:noncons}). In particular, asymptotically, one has the usual conservation equation for FLRW spacetimes in general relativity
\be
\dot{\rho}+3 H(\rho+p)=0, \label{eq:consnor}
\ee
where $\rho$ and $p$ are the total energy density and pressure. Hence, from the assumption that one has $\Lambda = T_{anomaly} = \mathcal{O}(H \Lambda_\textrm{QCD}^3)$ \cite{schutzhold:2002pl}, one deduces that asymptotically, one has \cite{borges:2005gr}
\be
\dot{\rho}_{m}+3 H (\rho_{m}+p_m)=-\dot{\Lambda}, \label{eq:lambda}
\ee
where $m$ here stands for the dust and radiation fields only.
The matter era solution in that case has been found in \cite{borges:2005gr}, and it indeed tends to the de Sitter solution for $t \rightarrow \infty$.

Hence, the de Sitter solution is indeed an asymptotic solution of entangled relativity, provided that the contribution to the on-shell Lagrangian from the trace anomaly is 
\be
\mathcal{L}_{anomaly}  = T_{anomaly} \propto H,
\ee
and $\L_{other} \sim 0$ in Eq. (\ref{eq:sumL}). Indeed, with this assumption,
\be
\phi \rightarrow \textrm{constant} \qquad \textrm{for $t \rightarrow \infty$},
\ee
follows from Eq. (\ref{eq:friction})---which starts to be valid (at least) at the begining of the matter era---from which one gets Eq. (\ref{eq:lambda}) asymptotically, such that \cite{borges:2005gr}
\be
\left(\rho_{dust} ,\rho_{radiation}, p_{radiation} \right) \rightarrow 0 \qquad \textrm{ for $t \rightarrow \infty$},
\ee
and
\be
a(t) \rightarrow  C e^{H_\infty t} \qquad \textrm{ for $t \rightarrow \infty$},
\ee
where $C$ is a constant of integration, whereas $H_\infty$ is the asymptotic constant value of the Hubble parameter. 

\subsection{Discussions}

The details of what happens during the transition---that is, before the variation of the scalar-field can safely be neglected---have yet to be studied. But let us stress again that the convergence of the scalar-field toward a constant simply follows from Eq. (\ref{eq:friction}), which happens for any combination of matter fields that satisfy $\L_m \sim T$ on-shell. 

Otherwise, the effects of perturbations will also have to be studied in order to further characterize the viability of this proposal---like for instance in \cite{basilakos:2009pr}, for the case that assumes the field equations of general relativity.
It may be interesting to focus on fields that do not satisfy $\L_m = T$---such as magnetic fields, which are such that $\L_{magnetic} \propto B^2 \neq T$---since those are the fields that should behave the most differently from their behaviors in general relativity.

\section{Summary}

In summary, in entangled relativity, the apparent current acceleration of the expansion of the universe might be explained by a coupling between the gravitational scalar degree of freedom of the theory to the quantum trace anomaly of the self-interacting quantum fields of the standard model in the cosmological space-time. Indeed, such a coupling induces a freezing of the scalar degree of freedom such that the phenomenology of the theory asymptotically reduces to the one of general relativity, while at the same time, it has been argued that the contribution of the QCD trace anomaly would have the correct amplitude in order to replace the cosmological constant as the driving force of the apparent expansion of the universe. As a corollary, most---if not all---of the phenomena related to the \textit{equivalence principle} would de facto be explained by the \textit{intrinsic decoupling} \cite{minazzoli:2013pr,minazzoli:2014pr,minazzoli:2016pr} of the scalar-degree of freedom at the level of the field equation (\ref{eq:sceq}). Therefore, it gives a plausible explanation to the acceleration of the expansion of the universe in the framework of entangled relativity that, notably, cannot afford to simply add a cosmological constant in its action---nor to do anything that would add a source to the extra degree of freedom---since the theory has to satisfy the tight constraints on the various manifestations of the equivalence principle.

On the opposite, the early times behavior of entangled relativity should be significantly different from the one of general relativity \cite{minazzoli:2018pr}. It would notably strongly depend on the on-shell value of $\mathcal{L}_m$ during the various phases of the early universe. This has yet to be explored.

\ack The author acknowledges support from the \textit{Fondation des fr\`eres Louis et Max Principale}, and thanks two anonymous referees for contributing to the improvement of the manuscript.

\section*{References}
\bibliographystyle{iopart-num}
\bibliography{ER_desitter}

\end{document}